\documentclass{article}
\usepackage{ijcai22}

\usepackage{times}

\usepackage{soul}
\usepackage{url}
\usepackage[hidelinks]{hyperref}
\usepackage[utf8]{inputenc}
\usepackage[small]{caption}
\usepackage{graphicx}
\usepackage{amsmath}
\usepackage{amsthm}
\usepackage{booktabs}
\usepackage{algorithm}
\usepackage{algorithmic}
\usepackage{tikz}
\usepackage{float}

\DeclareMathOperator*{\argmin}{\arg\!\,min}

\title{Learn 2 Rage: Experiencing The Emotional Roller Coaster That Is Reinforcement Learning}

\author{
Lachlan Mares 
\footnote{Contact Author}\and
Stefan Podgorski 
\And
Ian Reid
\affiliations
Australian Institute of Machine Learning, University of Adelaide\\
$^3$University of Adelaide\\
\emails
\{lachlan.mares, stefan.podgorski, ian.reid\}@adelaide.edu.au
}

\begin{document}
@Article{info11020125,
    AUTHOR = {Buslaev, Alexander and Iglovikov, Vladimir I. and Khvedchenya, Eugene and Parinov, Alex and Druzhinin, Mikhail and Kalinin, Alexandr A.},
    TITLE = {Albumentations: Fast and Flexible Image Augmentations},
    JOURNAL = {Information},
    VOLUME = {11},
    YEAR = {2020},
    NUMBER = {2},
    ARTICLE-NUMBER = {125},
    URL = {https://www.mdpi.com/2078-2489/11/2/125},
    ISSN = {2078-2489},
    DOI = {10.3390/info11020125}
}

@article{misra2019mish,
  title={Mish: A self regularized non-monotonic neural activation function},
  author={Misra, Diganta},
  journal={arXiv preprint arXiv:1908.08681},
  year={2019}
}

@article{Canny86a,
  added-at = {2017-05-10T00:00:00.000+0200},
  author = {Canny, John F.},
  biburl = {https://www.bibsonomy.org/bibtex/2fe58ad6db66de34cc55eb382f73946e4/dblp},
  ee = {http://doi.ieeecomputersociety.org/10.1109/TPAMI.1986.4767851},
  interhash = {9f13f83fc4620070b5e0365ba6d6acfb},
  intrahash = {fe58ad6db66de34cc55eb382f73946e4},
  journal = {IEEE Trans. Pattern Anal. Mach. Intell.},
  keywords = {dblp},
  number = 6,
  pages = {679-698},
  timestamp = {2017-05-11T11:39:17.000+0200},
  title = {A Computational Approach to Edge Detection.},
  url = {http://dblp.uni-trier.de/db/journals/pami/pami8.html#Canny86a},
  volume = 8,
  year = 1986
}

@inproceedings{NEURIPS2021_ba3c5fe1,
 author = {Shang, Wenling and Wang, Xiaofei and Srinivas, Aravind and Rajeswaran, Aravind and Gao, Yang and Abbeel, Pieter and Laskin, Misha},
 booktitle = {Advances in Neural Information Processing Systems},
 editor = {M. Ranzato and A. Beygelzimer and Y. Dauphin and P.S. Liang and J. Wortman Vaughan},
 pages = {22171--22183},
 publisher = {Curran Associates, Inc.},
 title = {Reinforcement Learning with Latent Flow},
 url = {https://proceedings.neurips.cc/paper/2021/file/ba3c5fe1d6d6708b5bffaeb6942b7e04-Paper.pdf},
 volume = {34},
 year = {2021}
}

@inproceedings{herman2021learn,
            title={Learn-to-Race: A Multimodal Control Environment for Autonomous Racing},
            author={Herman, James and Francis, Jonathan and Ganju, Siddha and Chen, Bingqing and Koul, Anirudh and Gupta, Abhinav and Skabelkin, Alexey and Zhukov, Ivan and Kumskoy, Max and Nyberg, Eric},
            booktitle={Proceedings of the IEEE/CVF International Conference on Computer Vision},
            pages={9793--9802},
            year={2021}
          }
          
@InProceedings{farneback03,
author="Farneb{\"a}ck, Gunnar",
editor="Bigun, Josef
and Gustavsson, Tomas",
title="Two-Frame Motion Estimation Based on Polynomial Expansion",
booktitle="Image Analysis",
year="2003",
publisher="Springer Berlin Heidelberg",
address="Berlin, Heidelberg",
pages="363--370",
abstract="This paper presents a novel two-frame motion estimation algorithm. The first step is to approximate each neighborhood of both frames by quadratic polynomials, which can be done efficiently using the polynomial expansion transform. From observing how an exact polynomial transforms under translation a method to estimate displacement fields from the polynomial expansion coefficients is derived and after a series of refinements leads to a robust algorithm. Evaluation on the Yosemite sequence shows good results.",
isbn="978-3-540-45103-7"
}

@inproceedings{HaarnojaZAL18,
  added-at = {2019-04-03T00:00:00.000+0200},
  author = {Haarnoja, Tuomas and Zhou, Aurick and Abbeel, Pieter and Levine, Sergey},
  biburl = {https://www.bibsonomy.org/bibtex/259e29cb876ddbd335ba7a1455d7e0a64/dblp},
  booktitle = {ICML},
  editor = {Dy, Jennifer G. and Krause, Andreas},
  ee = {http://proceedings.mlr.press/v80/haarnoja18b.html},
  interhash = {132a5861433d21c983754eb11e451022},
  intrahash = {59e29cb876ddbd335ba7a1455d7e0a64},
  keywords = {dblp},
  pages = {1856-1865},
  publisher = {PMLR},
  series = {Proceedings of Machine Learning Research},
  timestamp = {2019-04-04T11:43:21.000+0200},
  title = {Soft Actor-Critic: Off-Policy Maximum Entropy Deep Reinforcement Learning with a Stochastic Actor.},
  url = {http://dblp.uni-trier.de/db/conf/icml/icml2018.html#HaarnojaZAL18},
  volume = 80,
  year = 2018
}
   
@misc{wandb,
title = {Experiment Tracking with Weights and Biases},
year = {2020},
note = {Software available from wandb.com},
url={https://www.wandb.com/},
author = {Biewald, Lukas},
}

@misc{cosine,
  doi = {10.48550/ARXIV.1608.03983},
  url = {https://arxiv.org/abs/1608.03983},
  author = {Loshchilov, Ilya and Hutter, Frank},
  keywords = {Machine Learning (cs.LG), Neural and Evolutionary Computing (cs.NE), Optimization and Control (math.OC), FOS: Computer and information sciences, FOS: Computer and information sciences, FOS: Mathematics, FOS: Mathematics},
  title = {SGDR: Stochastic Gradient Descent with Warm Restarts},
  publisher = {arXiv},
  year = {2016},
  copyright = {arXiv.org perpetual, non-exclusive license}
}

@misc{hankyul2,
  author = {Han},
  title = {EfficientNetV2-pytorch},
  year = {2022},
  publisher = {GitHub},
  journal = {GitHub repository},
  howpublished = {\url{https://github.com/hankyul2/EfficientNetV2-pytorch}},
}

@article{https://doi.org/10.48550/arxiv.2104.00298,
  doi = {10.48550/ARXIV.2104.00298},
  url = {https://arxiv.org/abs/2104.00298},
  journal = {ICML},
  year = {2021},
  author = {Tan, Mingxing and Le, Quoc V.},
  keywords = {Computer Vision and Pattern Recognition (cs.CV), FOS: Computer and information sciences, FOS: Computer and information sciences},
  title = {EfficientNetV2: Smaller Models and Faster Training},
  publisher = {arXiv},
}

@misc{fpn,
  doi = {10.48550/ARXIV.1612.03144},
  url = {https://arxiv.org/abs/1612.03144},
  author = {Lin, Tsung-Yi and Dollár, Piotr and Girshick, Ross and He, Kaiming and Hariharan, Bharath and Belongie, Serge},
  keywords = {Computer Vision and Pattern Recognition (cs.CV), FOS: Computer and information sciences, FOS: Computer and information sciences},
  title = {Feature Pyramid Networks for Object Detection},
  publisher = {arXiv},
  year = {2016},
  copyright = {arXiv.org perpetual, non-exclusive license}
}

@misc{AIcrowd,
  author = {AIcrowd},
  title = {Crowdsourcing AI to Solve Real-World Problems},
  howpublished = {\url{https://www.aicrowd.com/}},
}

@misc{Arrival,
  author = {Arrival},
  title = {Explore Arrival products and technologies},
  howpublished = {\url{https://arrival.com/world/en}},
}

@article{stable-baselines3,
  author  = {Antonin Raffin and Ashley Hill and Adam Gleave and Anssi Kanervisto and Maximilian Ernestus and Noah Dormann},
  title   = {Stable-Baselines3: Reliable Reinforcement Learning Implementations},
  journal = {Journal of Machine Learning Research},
  year    = {2021},
  volume  = {22},
  number  = {268},
  pages   = {1-8},
  url     = {http://jmlr.org/papers/v22/20-1364.html}
}

@incollection{NEURIPS2019_9015,
title = {PyTorch: An Imperative Style, High-Performance Deep Learning Library},
author = {Paszke, Adam and Gross, Sam and Massa, Francisco and Lerer, Adam and Bradbury, James and Chanan, Gregory and Killeen, Trevor and Lin, Zeming and Gimelshein, Natalia and Antiga, Luca and Desmaison, Alban and Kopf, Andreas and Yang, Edward and DeVito, Zachary and Raison, Martin and Tejani, Alykhan and Chilamkurthy, Sasank and Steiner, Benoit and Fang, Lu and Bai, Junjie and Chintala, Soumith},
booktitle = {Advances in Neural Information Processing Systems 32},
pages = {8024--8035},
year = {2019},
publisher = {Curran Associates, Inc.},
url = {http://papers.neurips.cc/paper/9015-pytorch-an-imperative-style-high-performance-deep-learning-library.pdf}
} 

@article{DBLP:journals/corr/abs-2008-07971,
  author    = {Florian Fuchs and
               Yunlong Song and
               Elia Kaufmann and
               Davide Scaramuzza and
               Peter D{\"{u}}rr},
  title     = {Super-Human Performance in Gran Turismo Sport Using Deep Reinforcement
               Learning},
  journal   = {CoRR},
  volume    = {abs/2008.07971},
  year      = {2020},
  url       = {https://arxiv.org/abs/2008.07971},
  eprinttype = {arXiv},
  eprint    = {2008.07971},
  timestamp = {Mon, 09 Nov 2020 15:07:12 +0100},
  biburl    = {https://dblp.org/rec/journals/corr/abs-2008-07971.bib},
  bibsource = {dblp computer science bibliography, https://dblp.org}
}

@article{xavier,
author = {Glorot, Xavier and Bengio, Y.},
year = {2010},
month = {01},
pages = {249-256},
title = {Understanding the difficulty of training deep feedforward neural networks},
volume = {9},
journal = {Journal of Machine Learning Research - Proceedings Track}
}

@article{
doi:10.1126/scirobotics.abg5810,
author = {Antonio Loquercio  and Elia Kaufmann  and René Ranftl  and Matthias Müller  and Vladlen Koltun  and Davide Scaramuzza },
title = {Learning high-speed flight in the wild},
journal = {Science Robotics},
volume = {6},
number = {59},
pages = {eabg5810},
year = {2021},
doi = {10.1126/scirobotics.abg5810},
URL = {https://www.science.org/doi/abs/10.1126/scirobotics.abg5810},
eprint = {https://www.science.org/doi/pdf/10.1126/scirobotics.abg5810},
abstract = {Deep Learning enables agile flight in challenging environments with onboard sensing and computation. Quadrotors are agile. Unlike most other machines, they can traverse extremely complex environments at high speeds. To date, only expert human pilots have been able to fully exploit their capabilities. Autonomous operation with onboard sensing and computation has been limited to low speeds. State-of-the-art methods generally separate the navigation problem into subtasks: sensing, mapping, and planning. Although this approach has proven successful at low speeds, the separation it builds upon can be problematic for high-speed navigation in cluttered environments. The subtasks are executed sequentially, leading to increased processing latency and a compounding of errors through the pipeline. Here, we propose an end-to-end approach that can autonomously fly quadrotors through complex natural and human-made environments at high speeds with purely onboard sensing and computation. The key principle is to directly map noisy sensory observations to collision-free trajectories in a receding-horizon fashion. This direct mapping drastically reduces processing latency and increases robustness to noisy and incomplete perception. The sensorimotor mapping is performed by a convolutional network that is trained exclusively in simulation via privileged learning: imitating an expert with access to privileged information. By simulating realistic sensor noise, our approach achieves zero-shot transfer from simulation to challenging real-world environments that were never experienced during training: dense forests, snow-covered terrain, derailed trains, and collapsed buildings. Our work demonstrates that end-to-end policies trained in simulation enable high-speed autonomous flight through challenging environments, outperforming traditional obstacle avoidance pipelines.}}

\maketitle

\begin{abstract}
This work presents the experiments and solution outline for our teams winning submission in the Learn To Race Autonomous Racing Virtual Challenge 2022 hosted by \cite{AIcrowd}. The objective of the Learn-to-Race competition is to push the boundary of autonomous technology, with a focus on achieving the safety benefits of autonomous driving. In the description the competition is framed as a reinforcement learning (RL) challenge. 

We focused our initial efforts on implementation of Soft Actor Critic (SAC) variants. Our goal was to learn non-trivial control of the race car exclusively from visual and geometric features, directly mapping pixels to control actions. We made suitable modifications to the default reward policy aiming to promote smooth steering and acceleration control.

The framework for the competition provided real time simulation, meaning a single episode (learning experience) is measured in minutes. Instead of pursuing parallelisation of episodes we opted to explore a more traditional approach in which the visual perception was processed (via learned operators) and fed into rule-based controllers. 

Such a system, while not as academically ``attractive'' as a pixels-to-actions approach, results in a system that requires less training, is more explainable, generalises better and is easily tuned and ultimately out-performed all other agents in the competition by a large margin.

\end{abstract}

\section{Introduction}
\label{sec:introduction}

As autonomous vehicle technology advances and becomes part of the new normal there is greater importance for autonomous vehicles to adhere to safety standards. Whether the settings is urban driving or high-speed racing similar principals apply. Racing demands that a vehicle operate on the edge of its physical limits at all times dramatically increasing operational risks. Safe operation becomes critical, especially when any slight infraction could lead to catastrophic failure. Given financial investments in racing vehicles is high, autonomous racing in a controlled simulated environment can serve as a proving ground for development of safe learning algorithms.

The Learn-to-Race competition was a simulator-based challenge in which the aim was to build an autonomous agent to control a virtual racing car around a track. Agents were given prior learning opportunities to drive around a specific track, and then limited exposure to a new, different track, for which the goal was to achieve the fastest time without incurring safety infractions.

This paper describes our solution to this problem which achieved first place in single camera, and fastest overall track time in the 2022 edition of the challenge. We discuss the key features of our solution, as well as providing details of an academically more appealing, but significantly less successful approach based more directly on reinforcement learning. We use the contrast in these approaches to infer some lessons about learning for driving and other complex control tasks.

\subsection{Learn To Race competition}
\label{subsubsec:learn_to_race_competition}

The Learn-to-Race competition provides contestants with a Gym-compliant framework that leverages a high-fidelity racing simulator developed by \cite{Arrival}.

\begin{figure}
    \centering
    \includegraphics[width=0.47\textwidth]{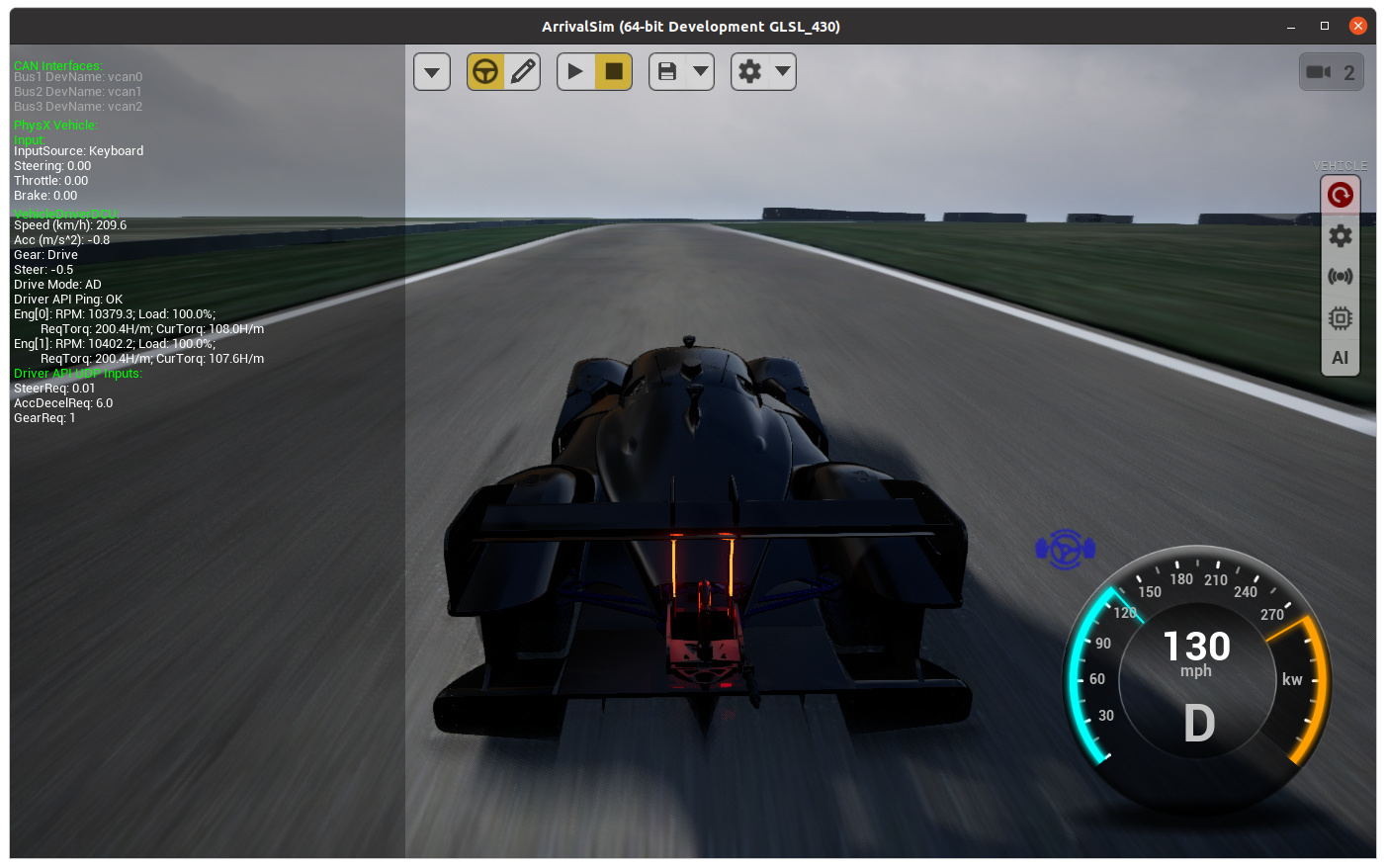}
    \label{fig:money_shot}
    \caption{Learn 2 Race simulator}
\end{figure}

It comprised two rounds. In round one participants developed and evaluated their agents on Thruxton Circuit, which was included with the Learn-to-Race environment. Due to the local availability of the round one evaluation track it was possible to perfectly fit an agent given the unbounded amount of practice time.  

In stage two agents were evaluated on an unseen track, the Vegas North Road Circuit. Evaluation consisted of a one hour practice period, then three laps of the circuit. The evaluation score considered safety infractions during training, together with safety infractions plus average speed for the additional three laps.

\subsubsection{Action space}
\label{subsubsec:action_space}
Agents execute actions within a simulated environment which models the dynamics of a vehicle and generates visual imagery from an agent-centric viewpoint. The action space comprises steering and acceleration control both with a continuous range of -1.0 to 1.0 representing full left to full right and maximum braking to maximum acceleration.

\subsubsection{Observation space}
\label{subsubsec:obervation_space}
\begin{figure}
    \centering
    \includegraphics[width=0.47\textwidth]{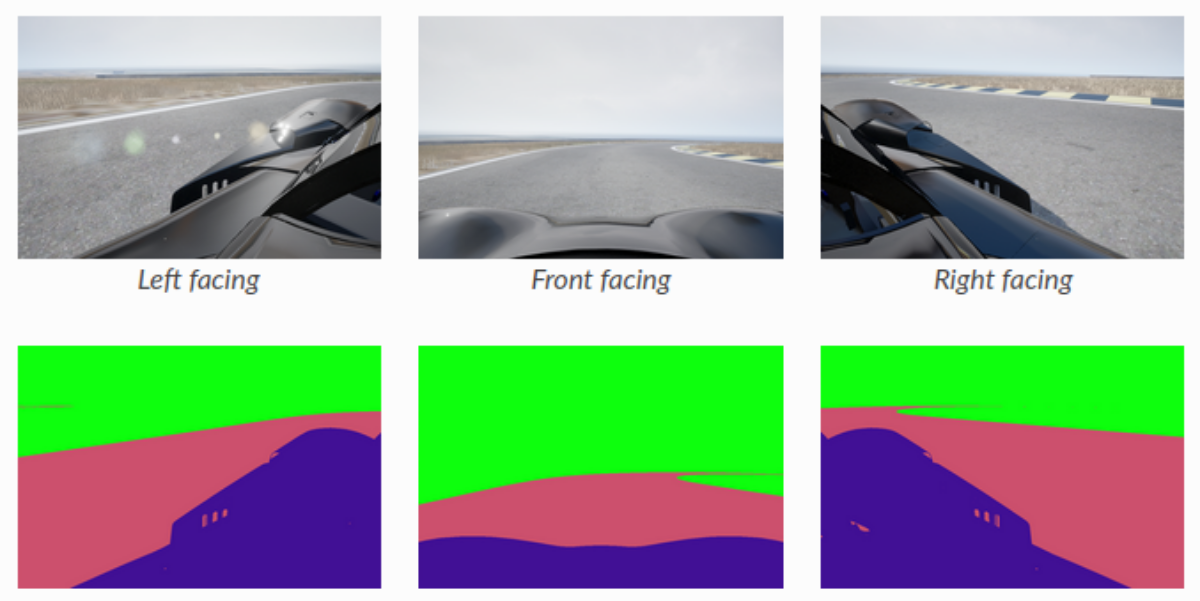}
    \caption{Observation images and segmentation masks}
    \label{fig:observation_space}
\end{figure}

The Learn-to-Race simulator provides competitors with multi-modal sensory inputs. The simulation environment provides for various observation data. During evaluation the only observation data available is the sequence of RGB images from a driver-centric perspective and the vehicle velocity. In addition, during development agents have access to semantic pixel labels (which designate where the road is in each image; see Figure \ref{fig:observation_space}), and the full vehicle state.

\subsubsection{Racetracks}
\label{subsubsec:racetracks}
The simulation environment included two real-world tracks. The first was Thruxton Circuit used for round one evaluation, modelled from the track at the Thruxton Motorsport Centre in the United Kingdom. The second is Anglesey National Circuit, located in Ty Croes, Anglesey, Wales.

For round two, the evaluation competition servers hosted an unseen track Vegas North Road, located at Las Vegas Motor Speedway in the United States.
\subsection{Our approach}
\label{sec:our_approach}

The objective of the Learn-to-Race competition is to push the boundary of autonomous technology, with a focus on achieving the safety benefits of autonomous driving.

We focused our initial efforts on implementing RL baselines \cite{stable-baselines3} such as Twin Delayed Deep Deterministic Policy Gradient (TD3) and Soft Actor Critic (SAC). In particular our goal was to learn non-trivial control of the race car exclusively from visual features, directly mapping pixels to control actions. We made suitable modifications to the default reward policy aiming to promote smooth steering and acceleration control. These experiments and results are outlined in Section \ref{sec:approach_1_reinforcement_learning}.

Nevertheless, the framework for the competition provided only for real time feedback, meaning a single episode (learning experience) is measured in minutes. Unless the training could be done via massive parallelisation of episodes (which we did not seek to do) the amount of training resources required was a significant limitation. 

As a baseline we decided also to explore a more traditional approach in which the visual percepts were processed (via learned operators) and fed into a rule-base controller. Such a system, while not as academically ``attractive'' as the pixels-to-actions approach, results in a system that requires less training, is more explainable, generalises better, is more easily tuned. Our system solution and experimental results are outlined in Section \ref{sec:approach_2_traditional_methods}.


We conclude the paper with a discussion of the relative strengths and weaknesses of the two approaches, lessons learned, and likely directions for future development.

\section{Approach 1: Reinforcement Learning}
\label{sec:approach_1_reinforcement_learning}

Our initial solution aims to use reinforcement learning (RL) to learn control commands based solely and directly from the RGB images ``observed'' by the agent. We term this approach a direct pixels-to-actions solution. While it is indeed possible to learn to race quickly and safely on a single track using this approach \cite{DBLP:journals/corr/abs-2008-07971}, a significant weakness of naive application of this approach is over-fitting to aspects of the imagery that are specific to the training track, and therefore a failure to generalise to a new track.  For this reason, and taking inspiration from \cite{doi:10.1126/scirobotics.abg5810} we pre-process the images in a number of ways, computing optical flow, scene segmentation and/or edge detection, as a means to train the system with features that generalise across domains and weather/lighting conditions. We further compress these features using a Variational Auto-Encoder (VAE) by only using the latent vector and discarding the decoder at run-time.

During the experiments the Soft Actor-Critic (SAC) \cite{HaarnojaZAL18} RL algorithm was used as the agent. A SAC agent comprises of two Multi-Layer Perceptrons (MLP), one is the actor and the other critic. The input to our SAC agents was obtained by passing the raw (multi-channel) image data through some pre-processing module/s to obtain a feature vector. The feature vector is a compressed representation containing relevant and important information about the environment.

In all our experiments the pre-processing module used was VAE. An Auto-Encoder (AE) is a deep-neural network designed to encode an image into a latent representation then reconstruct the image from the representation. A VAE was chosen in preference to an AE as the encoding distribution is regularised during training. We expect that this regularisation helps learn latent representations with better domain generalisation properties. We describe how these pre-processing elements are learned and used in the following sections.

\subsection{Continuous SAC}
\label{subsec:sac_continuous}

In this section we detail our experiments and trials using SAC with acceleration and steering action in continuous space using only VAE encoded RGB images for observations.

\subsubsection{SAC with FLARE}
\label{subsubsec:Sac_vanilla}
To improve the sample efficiency when training SAC Flow of Latents for Reinforcement Learning (FLARE) \cite{NEURIPS2021_ba3c5fe1} was integrated by taking the difference between the current and last timesteps latent feature vectors. SAC with FLARE (SAC-F) was used in all of the following RL experiments.

\subsubsection{SAC-F with Modified Reward Policy}
\label{subsubsec:sac_reward}
The default reward function for the competition was an adaption of the reward function presented in \cite{DBLP:journals/corr/abs-2008-07971}. The reward function has two components, a) track progression calculated based on centreline and b) avoiding track boundaries. We found during training the agent heavily favoured the track centre. Observed behaviour was the vehicle oscillating left and right trying to maintain track centre.

\begin{figure}
    \centering
    \includegraphics[width=0.4\textwidth]{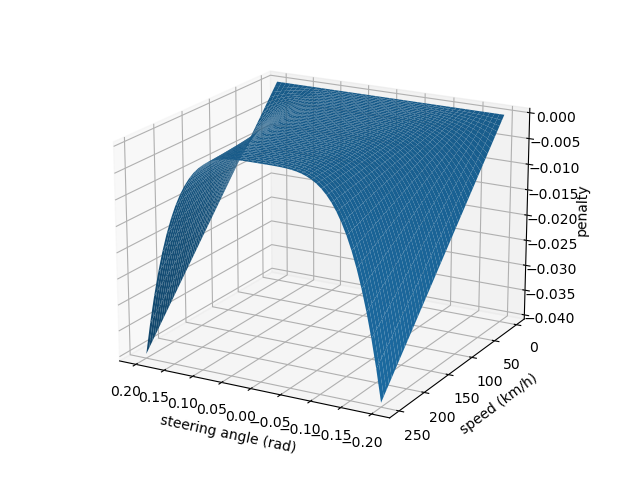}
    \caption{Steering penalty function for steering angle and speed.}
    \label{fig:sac_steering_penalty}
\end{figure}

Oscillating either side of track centre is inefficient, unstable and likely create unnecessary accidents and infractions during racing. With this in mind a modification to the reward was made to penalize the agent for making large steering changes. We formulate a steering penalty term:
\begin{equation*}
    \rho_{\text{steer}} (\alpha, s):=-\frac{s\alpha ^ 4}{10}
\end{equation*}
where $\alpha$ is the agent's speed and $s$ is the steering input, graphed in Figure \ref{fig:sac_steering_penalty}. This penalty was added to the reward at each time step with the intention of encouraging the agent to find a balance between smooth steering control and staying on the centerline. 

\begin{figure}
    \centering
    \includegraphics[width=0.4\textwidth]{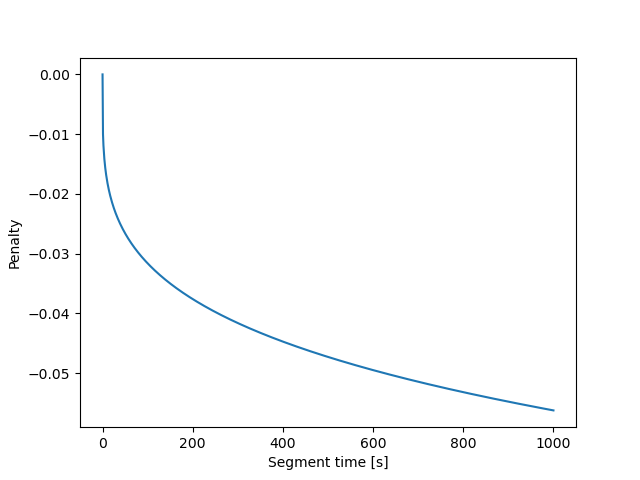}
    \caption{Segment penalty function for time agent is in a segment.}
    \label{fig:sac_segment_penalty}
\end{figure}

Learn-to-Race is a race, staying on track is necessary however the agent needs to drive as fast as possible. We found that using the default reward function did not result in fast track progression. To promote speed a penalty term was added to the reward function at each time step. This addition penalised the agent for the amount of time spent in each track segment. To implement this penalty a timer $t\in\mathcal{R}$ was started at the beginning of each track segment as the agent entered it. The penalty accumulated defined by:
\begin{equation*}
    \rho_{\text{segment}} (t):=-\frac{t^{1/4}}{100}
\end{equation*}
while the agent was inside the segment with $t$ resetting at the start of the next segment. The form of the penalty is shown in Figure \ref{fig:sac_segment_penalty}.

\subsubsection{Results}
\label{subsubsec:sac_continuous_results}

\begin{figure}
    \centering
    \includegraphics[width=0.47\textwidth]{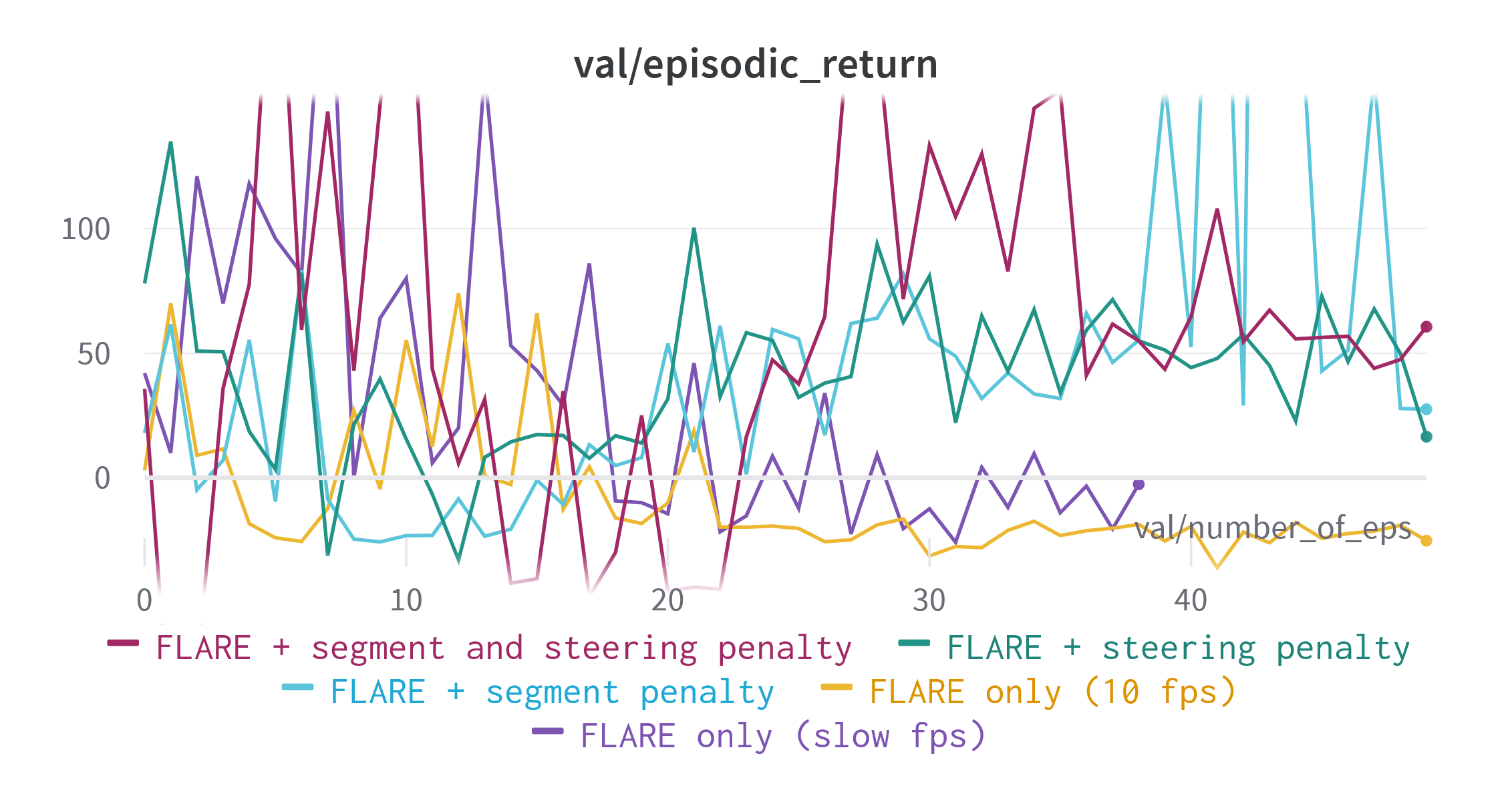}
    \caption{Cumulative reward for validation runs using the modifications to SAC.}
    \label{fig:sac_cumulative_results}
\end{figure}

Overall the performance of SAC-F with additional penalty terms was poor giving no successful completion of the track without safety infractions for any of the modifications.
The modifications to the reward functions showed promise for improving the validation episodic return. The small affect the additional penalty had on the cumulative reward shown in Figure \ref{fig:sac_cumulative_results} was seen after around 20 validations where the SAC-F agent trained with the penalties consistently returns positive reward.

\subsection{VAE with Optical-Flow and Image Edges}
\label{subsec:vae}
The purpose of the VAE is to transform the input image feature space to a lower dimensional latent space. The latent space is a compressed version of the input without unnecessary information. The VAE used in the previous methods were training on a data set of RGB images that were collected from the vehicles camera during manual control on the tracks that were available to entrants during the competition.

However, the final stage of the Learn-to-Race competition had entrants competing on an unseen track whose image distribution would not be the same as the training image distribution. The mismatch in the image distributions would cause poor generalisation due to over-fitting on the data set during training.

Therefore, the input images need to be pre-processed to extract features that describe the geometry of the observation for any track irrespective of specific image pixel values recorded by the camera. The new pre-processed images can then be used as the VAE input to produce abstract latent vectors for the agent.

The following methods of image pre-processing for abstract latent vectors were trialled: Binary segmentation masks (discussed in Section \ref{subsec:segmentation}), RGB representations of optical flow and, edge images.

Specific network details for the modified VAE can be found in Appendix \ref{subsec:variational_autoencoder_architecture_detail}.

\subsubsection{Optical Flow}
\label{subsubsec:optical_flow}
Optical flow was used to add information regarding the agents movement and to add geometry. The optical flow was generated using the Farneb\"ack method \cite{farneback03} and then converted into a three-channel RGB image.

\subsubsection{Edge Images}
\label{subsubsec:edge_images}
The images edges were used to add more detailed geometry information about the track which would be transferable by stripping away the rendered textures and leaving only the simulators geometry behind. The image edges were produced using two methods: Canny edge detection \cite{Canny86a} and edge detection using the S\"obel-Feldman operator.

\subsubsection{Custom VAE with Combined Inputs}
\label{subsubsec:custom_vae_with_combined_inputs}
The standard VAE used as a baseline in the Learn-to-Race competition needed to be modified and retrained to fit the new distributions formed by the abstract observations and their combinations. The baseline VAE was modified by increasing the depth of the network with an additional convolutional block at each layer, adding residual connections to the encoder and decoder convolutional blocks, using Mish activations \cite{misra2019mish} instead of ReLU, adding batch normalization to each convolutional block and initialising the VAE's weights using Xavier initialisation \cite{xavier}. No residual connections bypassed the bottleneck. The bottleneck connection required its output to be normalized using batch normalization for stability during and training.1st ICML Workshop on Safe Learning for Autonomous Driving (SL4AD)

\begin{figure*}
    \centering
    \begin{tabular}{ccc}
    \includegraphics[width=0.33\textwidth,height=0.2\textwidth]{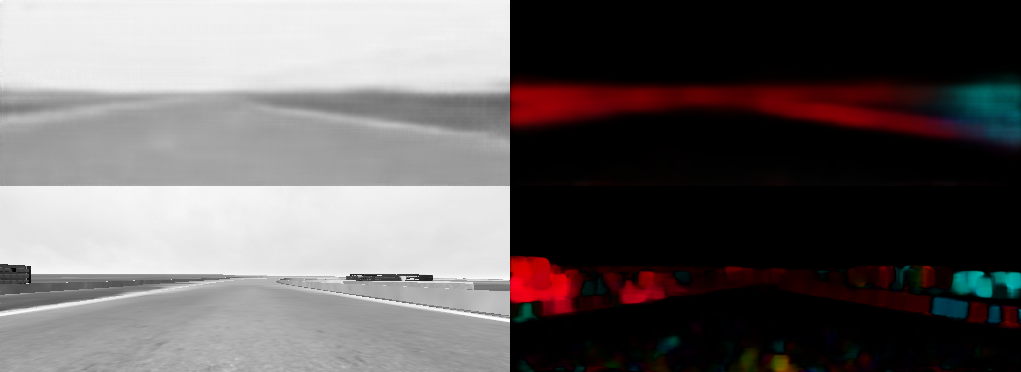} &
    \includegraphics[width=0.33\textwidth,height=0.2\textwidth]{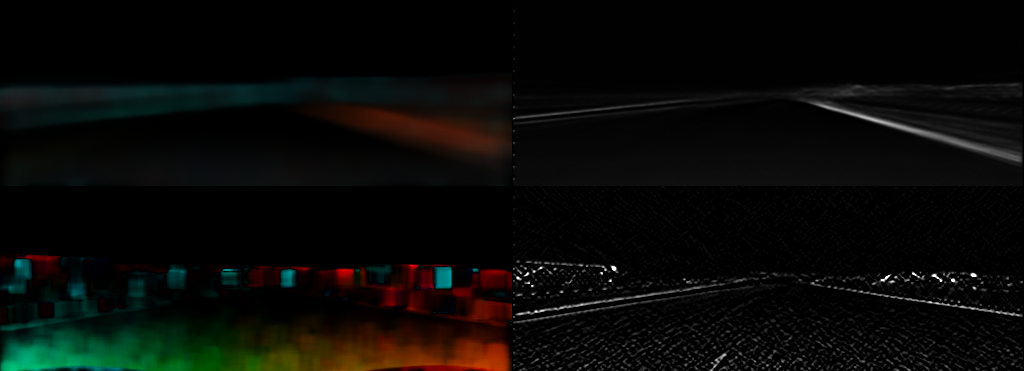} &
    \includegraphics[width=0.33\textwidth,height=0.2\textwidth]{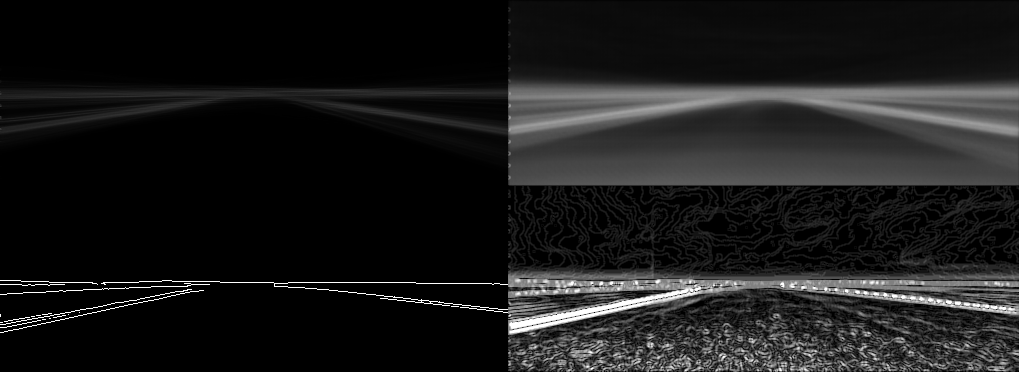} \\
    (a) & (b) & (c)    
    \end{tabular}
    \caption{Reconstruction results: Top row shows the reconstruction and bottom row is the ground truth. (a) gray scale and RGB flow (b) S\"obel edge images and RGB flow (c) S\"obel edge images and Canny edge images}
    \label{fig:reconstructions}
\end{figure*}

The modified abstract VAEs were trained on a NVIDIA 2080ti with cosine annealing \cite{cosine} for 1000 epochs starting with an initial learning rate of 0.001. A number of variations of input abstraction channels were trained which were combined to add information for the latent space. The variations we tested were Canny edges only, then gray scale image, see Figure \ref{fig:reconstructions}, each combined with the RGB optical flow.

\subsubsection{SAC-F Using Abstracted Latents}
\begin{figure*}
    \centering
    \begin{tabular}{ccc}
    \includegraphics[width=0.3\textwidth]{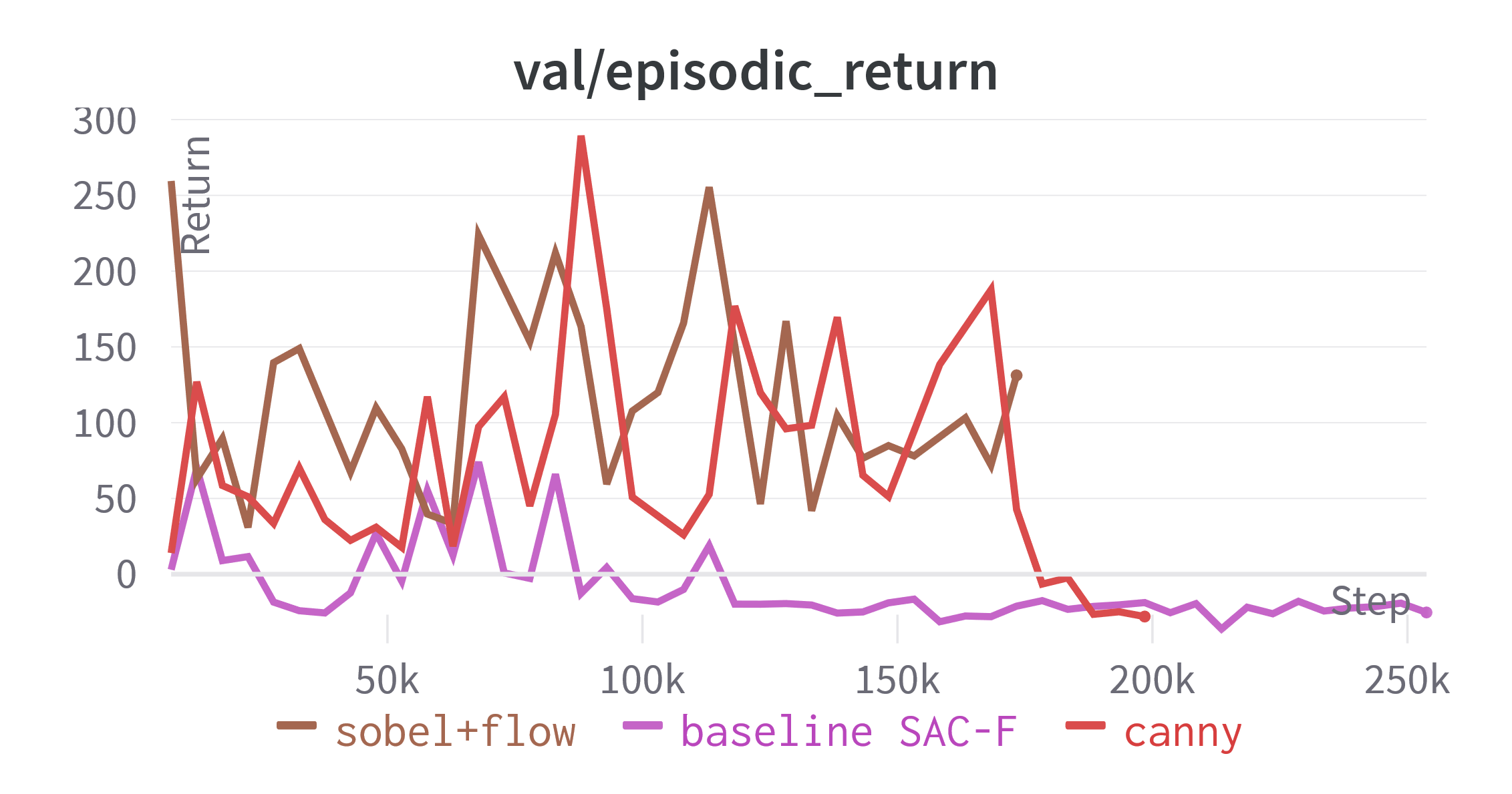}\label{fig:latents_return} &
    \includegraphics[width=0.3\textwidth]{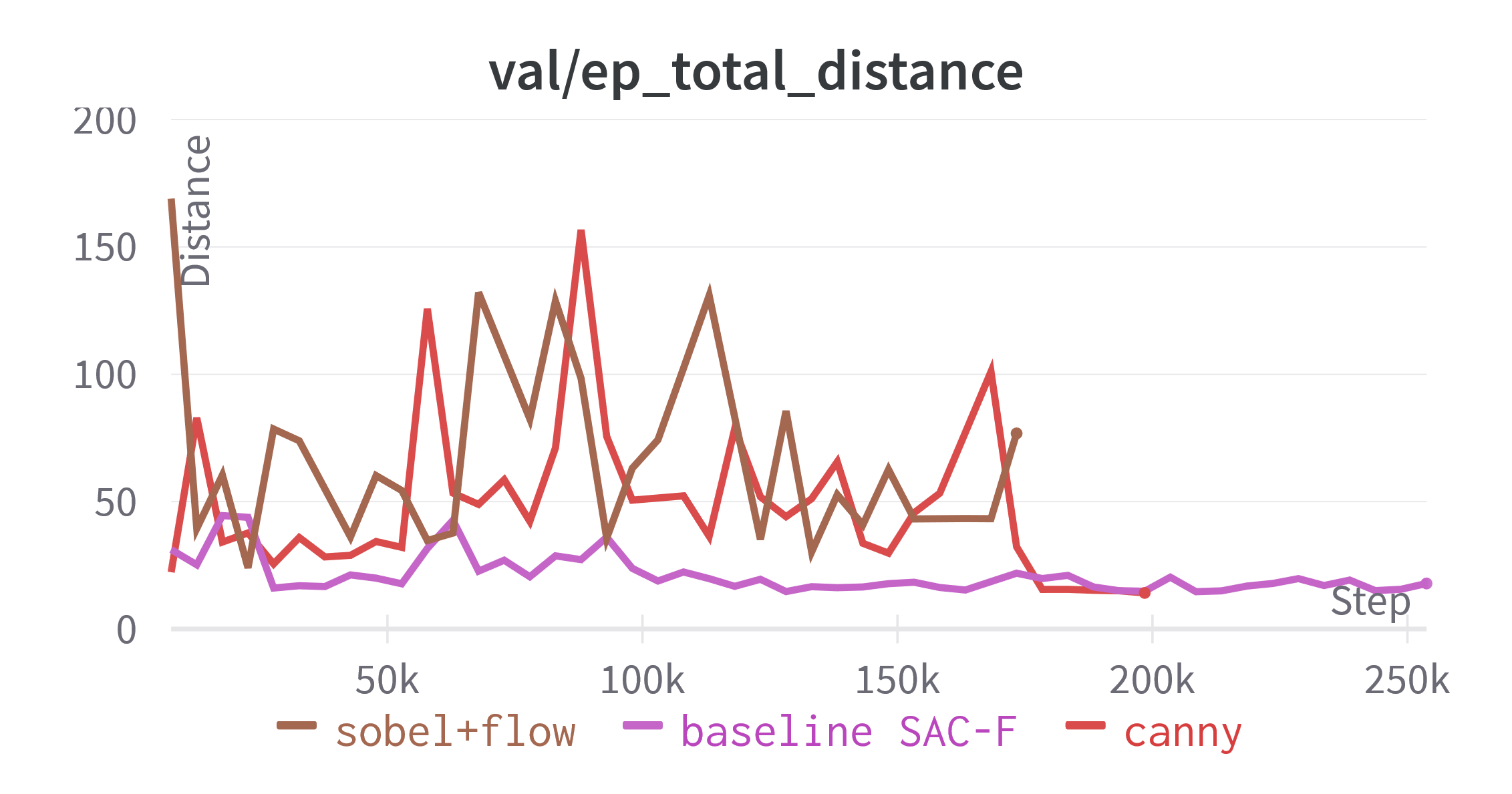}\label{fig:latents_dist} &
    \includegraphics[width=0.3\textwidth]{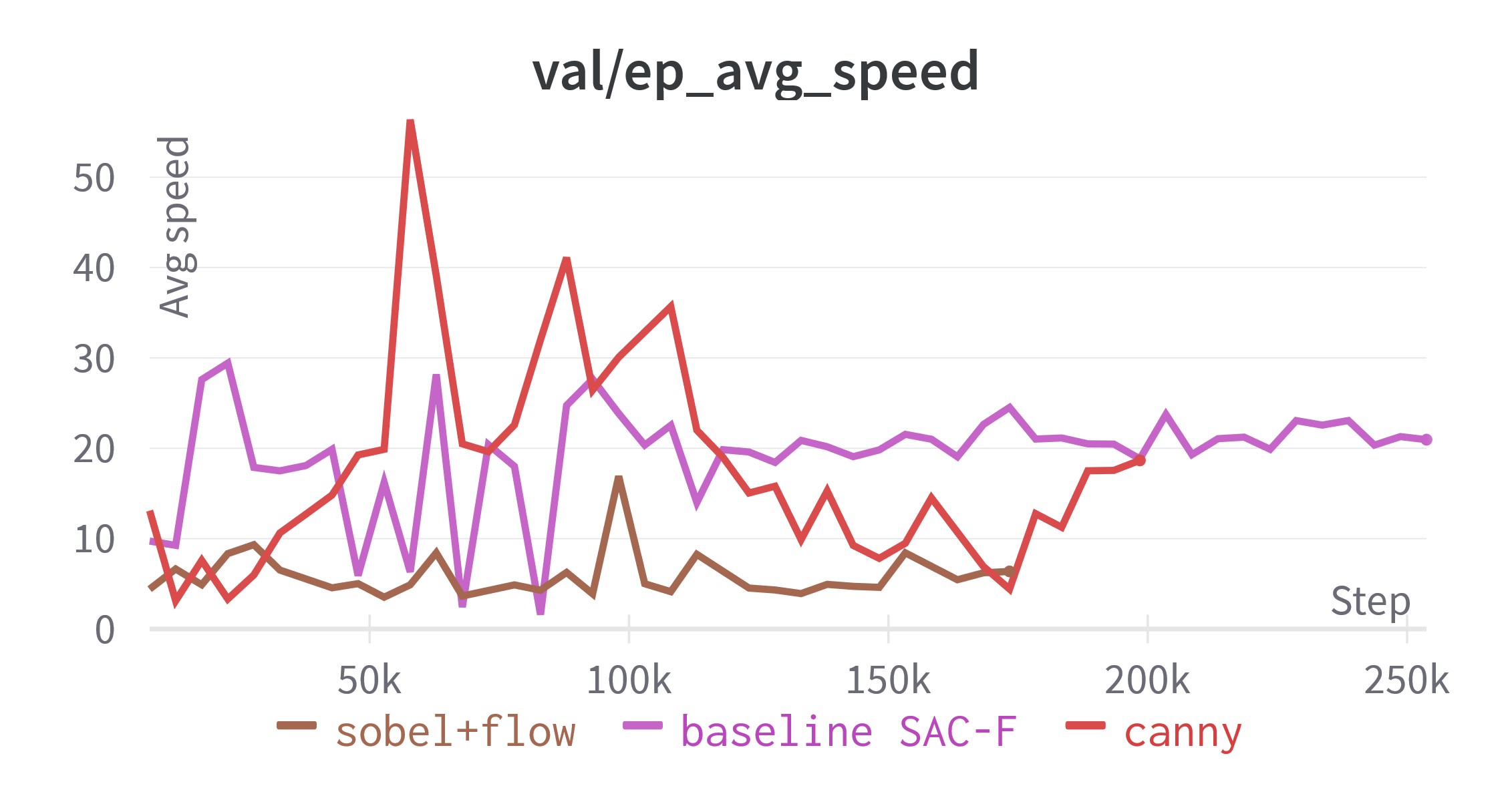}\label{fig:latents_speed} \\
    (a) & (b) & (c)    
    \end{tabular}
    \caption{Episodic return for the each of the trialled VAEs. (a) Episodic return is generally higher for the abstract latents than the  baseline method. (b) total distance travelled in the validation episode was in general higher than the baseline method and (c) average speed was in general higher for the baseline method than the abstract latents.}
    \label{fig:rl_results}
\end{figure*}

Three methods were chosen to trial, the baseline RGB VAE with FLARE \cite{info11020125}, RGB optical flow over the previous and current frame with an additional S\"obel image edge channel, and Canny edges only. The combined segmentation masks and RGB optical flow did not appear to reconstruct a detailed enough image and was not trialled which indicate that the latent vectors did not contain enough information. The methods were trialled for 10-20 hours on a NVIDIA 2080ti with a frame rate of ~5 fps. Overall, the use of the abstracted latents marginally improved the total distance travelled (Figure \ref{fig:rl_results} (a)) and episodic return (Figure \ref{fig:rl_results} (b)) during validation on Thruxton when compared to the baseline method. However, the average speed (Figure \ref{fig:rl_results}(c)) was negatively affected by the addition of the abstract latents. 

\subsection{RL approach summary}
\label{subsec:rl_approach_summary}

Due to the general applicability of model free RL algorithms should be the ideal candidate for complex control tasks such as autonomous racing applications. The ability to learn from experience, adapt and find unique policies in theory should be able to out-perform any rule based solution. 

The issue we faced during the competition is the optimal policy is learnt from the environment through extensive exploration, this process however takes thousands of experiences from thousands of environment steps costing time and computing resources that we were unwilling to commit given the one hour time constraint in round two. 
\section{Approach 2: Traditional methods}
\label{sec:approach_2_traditional_methods}
As a minimum baseline we also implemented a more traditional approach to autonomous agent control. This approach uses a combination of (learned) image processing to achieve situation awareness (track location, position, etc) and rule-based control.

In particular, we implemented two localisation modules: the first identified where the centreline of the road in current observation, as the primary sensor signal to influence steering angle. The second module identified which zone of the track the agent was currently traversing. This was used to inform the agent of upcoming acceleration and braking zones. Such zones can only be identified in the broader context of a memory of the whole track, since they often cannot be disambiguated using the immediate visual stimulus alone. Without positional awareness the acceleration controllers ability to safely regulate speed was constrained by the camera field of view.

Given the simulator provided segmentation masks a logical choice was to use a semantic segmentation network for track surface identification. The segmentation network accurately classified image pixels as either track or not-track. With observable pixels classified, key geometric properties of the observation could be calculated. The properties of the observation can then be passed to steering and acceleration controllers. 

\subsection{Semantic segmentation}
\label{subsec:segmentation}

This section discusses our semantic segmentation model used to extract the drivable area ahead of the agent. We first discuss the model architecture utilised for agent perception then discuss the offline training regime. Finally we present the fine tuning used by our agent during the one hour training period before final evaluation.

\subsubsection{Segmentation model architecture}
\label{subsubsec:segmentation_model_architecture}
The semantic segmentation model required a balance between low inference time and high accuracy. The architecture chosen was a custom PyTorch \cite{NEURIPS2019_9015} implementation consisting of an EfficientNet-V2-Small encoder paired with a Feature Pyramid Network (FPN) decoder.

EfficientNetV2 is a convolutional neural network that has been optimised to increase training speed and maximise parameter efficiency. To develop these models, the authors \cite{https://doi.org/10.48550/arxiv.2104.00298} use a combination of training-aware neural architecture search and scaling, to jointly optimize training speed. EfficientNetV2 utilises a new Fused-MBConv in conjunction with the MBConv present in EfficientNet. The EfficientNetV2 encoder used for the challenge was adapted from an existing PyTorch repository \cite{hankyul2}.

FPN is a feature extractor for segmentation networks proposed by \cite{fpn}. The FPN feature extractor generates multiple feature map layers (multi-scale feature maps) with higher quality information than the regular feature pyramid for object detection. 

Specific network details can be found in Appendix \ref{subsec:segmentation_network_architecture_detail}.

\subsubsection{Segmentation model training}
\label{sec:segmentation_model_training}
\begin{figure}
    \centering
    \includegraphics[width=0.4\textwidth]{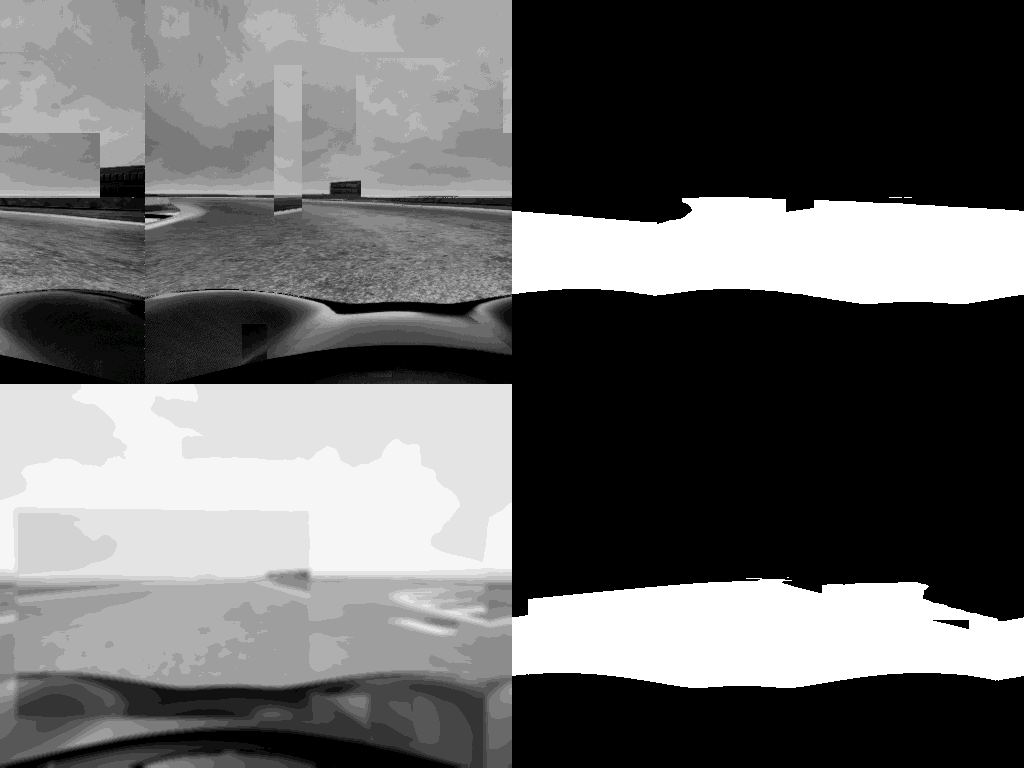}
    \caption{Augmentation Examples}
\label{fig:augmentations}
\end{figure}

The segmentation model was trained using 384x512 grey-scale images to reduce the expected domain gap between unobserved tracks. Extensive image augmentations were used to further extend the captured 25000+ training examples. Image and mask augmentation was primarily handled with the Albumentations library \cite{info11020125}. Augmentations included Horizontal-Flip, Blur, CLAHE, Posterize, Random Contrast and Random Brightness. Additional custom augmentations included axis roll and random cropping. See Figure~\ref{fig:augmentations} for some training samples.

The segmentation model was trained using a standard Adam optimiser and a custom implementation of Dice Loss:
\begin{equation*}
\text{Dice Loss} = 1 - \frac{2*TP}{2*TP+FP+FN}
\label{eq:diceloss}
\end{equation*}
\\
\subsubsection{Segmentation model training during evaluation}
\label{subsec:segmentation_model_during_evaluation}
The Learn To Race Challenge allowed participants to see the test track for one hour prior to evaluation. We utilise this time to fine tune the segmentation model to ensure high accuracy predictions. During this training period our agent conservatively drives around the track inferring the drivable area. The loss between the inferred prediction and the provided ground truth is then calculated. If the loss is higher than a tunable threshold it is saved to be used for fine tuning later in the evaluation period. It was found that to ensure stability when training images and masks from Thruxton and Anglesey also needed to be included in the fine tune training.

\subsection{Steering controller}
\label{sec:vehicle_steering_control}
\begin{figure}
    \centering
    \includegraphics[width=0.47\textwidth]{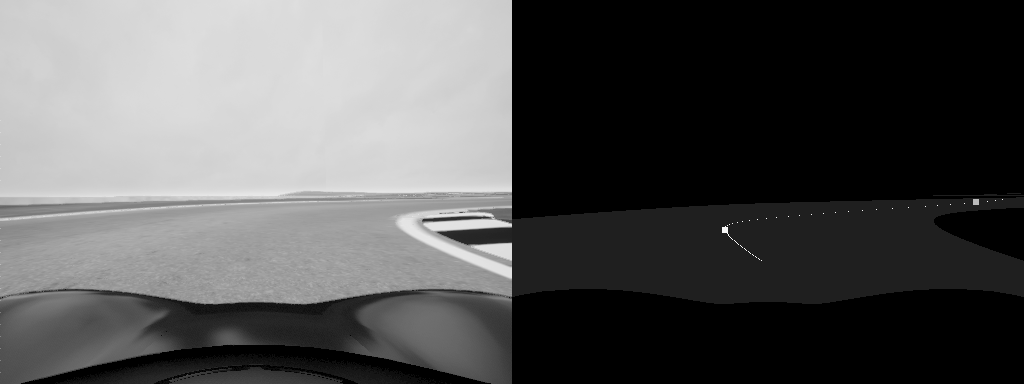}
    \caption{Upper/Lower bounds of usable centreline pixels}
    \label{fig:image_steering_overlay}
\end{figure}

The steering and acceleration controllers both rely on predictions from the segmentation model. An estimate of the track centreline is calculated using the boundaries of the predicted road surface shown in Figure~\ref{fig:image_steering_overlay}. The steering angle was calculated in pixel coordinates rather than transforming to Cartesian coordinates. Based on the vehicles current velocity a selection of centreline pixels were used to calculate the relative angle $RA$ for the next environment step:

\begin{equation*}
     RA =  \frac{\arctan{(P_{j},(P_{i} - (w/2))) - \pi/2}}{\pi/2}
\label{eq:relative_steering}
\end{equation*}
where $P_{ij}$ is the centreline pixel indices, and $w$ is the image width.

The relative steering angle was fed into a Proportional Integral Differential (PID) controller.

%


\subsection{Acceleration controller}
\label{subsec:vehicle_acceleration_control}
To approximate future time-steps the track centreline pixels were split into discrete chunks. The chunk size was chosen so that it represented expected future vehicle positions in the observed image. The chunks were then used to compute an estimation of the track curvature at each future step. A Gaussian weighting parameter W was used to weight the future track curvature and current velocity:
\begin{equation*}
     \text{W} =  1 - \frac{1}{c} \sum_{j=1}^c \left(\exp\left( -\frac{1}{2}\left(\frac{x-\mu}{\sigma}\right)^{\!2}\,\right)\right)
\label{eq:weighting_parameter}
\end{equation*}
where $c$ is the number of chunks, $\mu$ is $c$ evenly distributed values between 0 and 1, $x$ is the current velocity as ratio of maximum velocity set-point, and $\sigma$ is the the spread of individual chunks.



Taking inspiration from human drivers the acceleration controller used a discrete set of actions. The simplification of the action space permitted the use of a simple Bang-Bang controller to output one of five states given the current velocity and $W$. Action space used values [-1, 0, 0.2, 0.4, 1], 0.2 and 0.4 were used if current steering angle exceeded 0.5 and 0.3 respectively.    

\subsection{Results without track localisation}
\label{subsec:vehicle_wothout_track_localisationl}
Work on track localisation started towards the end of round one of the Learn-to-Race competition. Round one submission could have certainly been improved with localisation, but given finishing in the top 10 granted access to Round two it was not pursued. The results for the three tracks pre-localisation are shown in Table \ref{tab:track_times_pre}

\begin{table}
    \centering
    \begin{tabular}{l c c| c c c}
        \hline
        Track  & $fps$ & local kph & $fps$ & eval kph \\
        \hline
        Vegas North Road   &     &        & 6.4 & 70.853 \\
        Anglesey National & 7.5 & 85.89  &     &        \\
        Thruxton         & 7.5 & 132.56 & 6.4 & 126.350 \\
        \hline
    \end{tabular}
    \caption{Track times without location classifier, $fps$ frames per second, local/eval (AIcrowd) refers to simulator location}
\label{tab:track_times_pre}
\end{table}

\subsection{Track localisation with classification models}
\label{subsec:track_localisation_with_classification_models}
Without localisation the vehicles speed is limited by the track in which it can observe. Removing the threat of a tight corner appearing unexpectedly while traveling at top speed allows the agent to travel safely while also obtaining higher average speeds. 

Experiments were conducted with three different implementations of location classification, all using a neural network. The first utilises the track segments provided alongside the simulator, the second uses an arbitrary number of segments spread out around the map, and the third requires hand labeling of distinct zones.


\subsubsection{Classification model architecture}
\label{subsubsec:classification_model_architecture}
The classification model used during experimentation was a two layer MLP. Input to the classification model was a feature vector extracted from the semantic segmentation model. The architecture, dimensions and activation functions are listed in Table~\ref{tab:classifier_layers}.

\begin{table}
    \centering
    \begin{tabular}{l| c c c}
        \hline
        Layer     & Input Size     & Output Size   & Activation \\
        \hline
        Input     & feature\_size  & feature\_size & Tanh \\
        Hidden 1  & feature\_size  & 512           & ReLU \\
        Hidden 2  & 512            & 256           & ReLU \\
        Output    & 256            & classes       & Softmax \\
        \hline
    \end{tabular}
    \caption{MLP Layers used for location classification}
\label{tab:classifier_layers}
\end{table}

Specific network details can be found in Appendix \ref{subsec:classifier_network_architecture_detail}.

\subsubsection{Experiment 1: Environment track segment}
\label{subsubsec:experiment_1}
The goal of this experiment was to train a classification model to recognise the current track segment provided by the simulator. The provided code base which runs experiments tracks the current segment. Probing this code at run time allowed us to label samples with the current segment. The model was trained using 5000 samples spanning the 10 segments of Thruxton Circuit. The model achieved an accuracy of 99.76\% on the 5000 samples collected. The errors observed during evaluation occurred at the boundary where two track segments met. 

This classifier was not utilised in round two as knowing the vehicles location at such a coarse scale offered no benefit predicting upcoming difficult track sections.

\subsubsection{Experiment 2: N Sections}
\label{subsubsec:experiment_2}
The goal of this experiment was to train a classification model to recognise up to $N$ discrete track sections. Track sections were defined by evenly selecting key-points along the tracks centreline, a JSON file containing centreline points $(x, y)$ was included with the simulator. The $N$ key-points $(key_x, key_y)$ were then used to divide training samples into sections:
\begin{equation*}
    \text{section} = \argmin\sqrt{(\text{key}_x - x)^2 + (\text{key}_y - y)^2}
\label{eq:track_sections}
\end{equation*}

The model was trained using 10000 samples spanning 40 sections. The model achieved an accuracy of 96.4\% on samples gathered from Thruxton circuit. The observed errors during evaluation were not confined to the boundaries where two track sections met as with Experiment 1. The observation errors were able to be corrected with a transition filter which constrained section changes to occur in sequence.

This approach was a viable option but was not utilised in round two due to a preference for the method outlined in Experiment 3.

\subsubsection{Experiment 3: Specific zone}
\label{subsubsec:experiment_3}
The goal of this experiment was to train a classification model to recognise specific track zones deemed difficult or required actions beyond recognition of sensors. Track zones were categorised as one of three discrete classes: 

\begin{itemize}
    \item 0, High speed areas, straights and exits from corners
    \item 1, Approaching moderate corners at mid velocity, tight corners at low velocity
    \item 2, Approaching moderate and tight corners such as chicanes and hairpins at high velocity
\label{item:track_zone_types}
\end{itemize}

\begin{figure}
    \centering
    \includegraphics[width=0.47\textwidth]{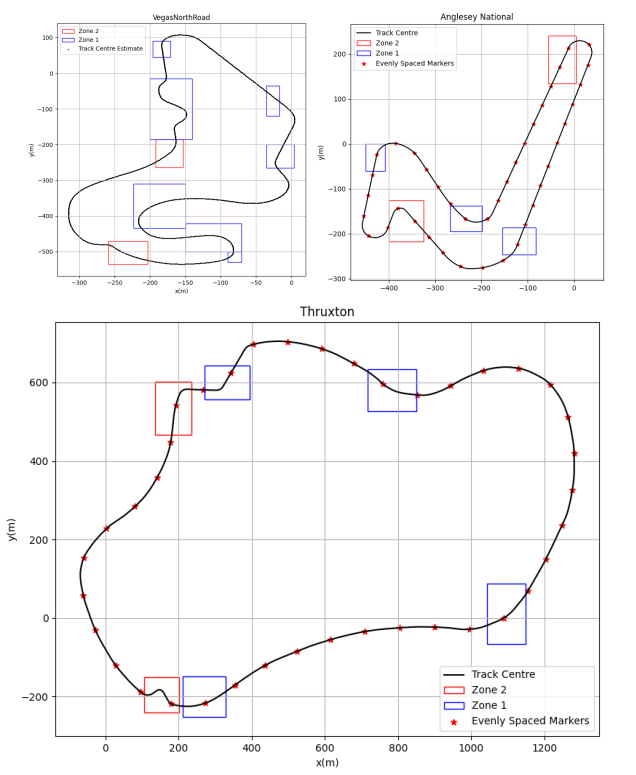}
    \caption{Zone boundaries for all three tracks}
    \label{fig:track_plots}
\end{figure}

Selecting track zones was a manual process based on intuition and experiments. 
For example a long straight followed by a chicane would is labelled as zone 2 as the agent will be approaching at high speed and need to slow down.
The defined zones were specified as boxes $(xmin, ymin, xmax, ymax)$ and labeled [0, 1, 2]. Setting zones for each track is an iterative process starting conservative then altering them based upon feedback from observation, evaluation logs and results. Experimentation was conducted on Thruxton and Anglesey National Circuits, zones are shown in Figure~\ref{fig:track_plots}.

The model was trained using 10000 combined samples, 5000 from both Thruxton and Anglesey National Circuits.
The model achieved an accuracy of 99.62\% during evaluation laps of Thruxton, and 99.69\% during evaluation laps of Anglesey National. 
As seen in Experiment 1 the observed errors during evaluation occurred where two track zones met. 
Out of the three experiments the specific zone classifier was the easiest to configure, had the best generalisation and was trainable within the one hour training window. 

This version of location classifier was utilised in the round two of Learn to Race, and was a major contributor to winning the competition.

\subsubsection{Classifier model training during evaluation}
\label{sec:classifier_model_training_during_evaluation}
To collect the samples required to train the localisation classifier the vehicle was set to follow the estimated centreline of the track at up to 10m/s. 
At each environment step the current zone was calculated using the zone boundaries shown in Figure \ref{fig:track_plots}, then the feature vector from the already fine tuned segmentation model and zone were saved.
A total of 4500 training samples were collected.
Since the majority of the collected samples are within zone 0 balancing of classes was required. We collect an additional 500 samples of zone 1 and 2 to combat this. 

No augmentations were applied to the feature vector during training and no train-validation split was used. The intention during training was to over-fit the model to the data. Adam optimiser and cross entropy loss were used for training as per aforementioned experiments. The model accuracy recorded for our best submission was 97.7\% after 150 epochs, the lower than expected accuracy led to slightly inconsistent zone predictions. Without local verification it is unclear how much effect the mis-classifications had on track times.

\subsubsection{Results with track localisation}
\label{sec:approach_2_summary}
Of the three classifiers Specific Zone was chosen for round 2. The work implementing the localisation proved to be crucial for posting the fastest lap time in round two and ultimately winning the Learn-to-Race competition. This implementation posted a 21.6kph improvement in average speed compared to initial submission. The results for the three tracks with localisation are shown in Table \ref{tab:track_times_post}.
 
\begin{table}
    \centering
    \begin{tabular}{l c c| c c}
        \hline
        Track  & $fps$ & local kph & $fps$ & eval kph \\
        \hline
        Vegas North Road   &     &        & 5.1 & 92.54 \\
        Anglesey National & 7.0 & 103.97 &     &       \\
        Thruxton         & 7.0 & 147.12 &     &       \\
        \hline
    \end{tabular}
    \caption{Track times with location classifier, $fps$ frames per second, local or evaluation (AIcrowd) refers to simulator location}
    \label{tab:track_times_post}
\end{table}

\section{Conclusion}
\label{sec:conclusion}

In this paper, we have presented the experiments and final approach taken in the Learn-to-Race challenge for safe autonomous driving 2022. 

In summary, the RL agents from our experiments were inefficient to train and offered no significant improvement compared to the provided Learn-to-Race baseline agents. The inability to effectively explore and sample the unobserved evaluation environments resulted in our agents exhibiting undesired behaviours that were prone to safety infractions.

The major benefit of our traditional approach was its ability to generalise, requiring minimal training time to perform well in unobserved environments. There are still a few key areas that can be explored to increase performance for future work these include a) streamlining of code to increase the observation and controller update rate, b) implement more sophisticated steering and acceleration controllers, and c) explore bootstrapping an RL agent with current approach essentially using RL as fine-tuning mechanism.


\bibliographystyle{named}
\bibliography{refs}

\begin{thebibliography}{}

\bibitem[\protect\citeauthoryear{AIcrowd}{}]{AIcrowd}
AIcrowd.
\newblock Crowdsourcing ai to solve real-world problems.
\newblock \url{https://www.aicrowd.com/}.

\bibitem[\protect\citeauthoryear{Arrival}{}]{Arrival}
Arrival.
\newblock Explore arrival products and technologies.
\newblock \url{https://arrival.com/world/en}.

\bibitem[\protect\citeauthoryear{Buslaev \bgroup \em et al.\egroup }{2020}]{info11020125}
Alexander Buslaev, Vladimir~I. Iglovikov, Eugene Khvedchenya, Alex Parinov, Mikhail Druzhinin, and Alexandr~A. Kalinin.
\newblock Albumentations: Fast and flexible image augmentations.
\newblock {\em Information}, 11(2), 2020.

\bibitem[\protect\citeauthoryear{Canny}{1986}]{Canny86a}
John~F. Canny.
\newblock A computational approach to edge detection.
\newblock {\em IEEE Trans. Pattern Anal. Mach. Intell.}, 8(6):679--698, 1986.

\bibitem[\protect\citeauthoryear{Farneb{\"a}ck}{2003}]{farneback03}
Gunnar Farneb{\"a}ck.
\newblock Two-frame motion estimation based on polynomial expansion.
\newblock In Josef Bigun and Tomas Gustavsson, editors, {\em Image Analysis}, pages 363--370, Berlin, Heidelberg, 2003. Springer Berlin Heidelberg.

\bibitem[\protect\citeauthoryear{Fuchs \bgroup \em et al.\egroup }{2020}]{DBLP:journals/corr/abs-2008-07971}
Florian Fuchs, Yunlong Song, Elia Kaufmann, Davide Scaramuzza, and Peter D{\"{u}}rr.
\newblock Super-human performance in gran turismo sport using deep reinforcement learning.
\newblock {\em CoRR}, abs/2008.07971, 2020.

\bibitem[\protect\citeauthoryear{Glorot and Bengio}{2010}]{xavier}
Xavier Glorot and Y.~Bengio.
\newblock Understanding the difficulty of training deep feedforward neural networks.
\newblock {\em Journal of Machine Learning Research - Proceedings Track}, 9:249--256, 01 2010.

\bibitem[\protect\citeauthoryear{Haarnoja \bgroup \em et al.\egroup }{2018}]{HaarnojaZAL18}
Tuomas Haarnoja, Aurick Zhou, Pieter Abbeel, and Sergey Levine.
\newblock Soft actor-critic: Off-policy maximum entropy deep reinforcement learning with a stochastic actor.
\newblock In Jennifer~G. Dy and Andreas Krause, editors, {\em ICML}, volume~80 of {\em Proceedings of Machine Learning Research}, pages 1856--1865. PMLR, 2018.

\bibitem[\protect\citeauthoryear{Han}{2022}]{hankyul2}
Han.
\newblock Efficientnetv2-pytorch.
\newblock \url{https://github.com/hankyul2/EfficientNetV2-pytorch}, 2022.

\bibitem[\protect\citeauthoryear{Lin \bgroup \em et al.\egroup }{2016}]{fpn}
Tsung-Yi Lin, Piotr Dollár, Ross Girshick, Kaiming He, Bharath Hariharan, and Serge Belongie.
\newblock Feature pyramid networks for object detection, 2016.

\bibitem[\protect\citeauthoryear{Loquercio \bgroup \em et al.\egroup }{2021}]{doi:10.1126/scirobotics.abg5810}
Antonio Loquercio, Elia Kaufmann, René Ranftl, Matthias Müller, Vladlen Koltun, and Davide Scaramuzza.
\newblock Learning high-speed flight in the wild.
\newblock {\em Science Robotics}, 6(59):eabg5810, 2021.

\bibitem[\protect\citeauthoryear{Loshchilov and Hutter}{2016}]{cosine}
Ilya Loshchilov and Frank Hutter.
\newblock Sgdr: Stochastic gradient descent with warm restarts, 2016.

\bibitem[\protect\citeauthoryear{Misra}{2019}]{misra2019mish}
Diganta Misra.
\newblock Mish: A self regularized non-monotonic neural activation function.
\newblock {\em arXiv preprint arXiv:1908.08681}, 2019.

\bibitem[\protect\citeauthoryear{Paszke \bgroup \em et al.\egroup }{2019}]{NEURIPS2019_9015}
Adam Paszke, Sam Gross, Francisco Massa, Adam Lerer, James Bradbury, Gregory Chanan, Trevor Killeen, Zeming Lin, Natalia Gimelshein, Luca Antiga, Alban Desmaison, Andreas Kopf, Edward Yang, Zachary DeVito, Martin Raison, Alykhan Tejani, Sasank Chilamkurthy, Benoit Steiner, Lu~Fang, Junjie Bai, and Soumith Chintala.
\newblock Pytorch: An imperative style, high-performance deep learning library.
\newblock In {\em Advances in Neural Information Processing Systems 32}, pages 8024--8035. Curran Associates, Inc., 2019.

\bibitem[\protect\citeauthoryear{Raffin \bgroup \em et al.\egroup }{2021}]{stable-baselines3}
Antonin Raffin, Ashley Hill, Adam Gleave, Anssi Kanervisto, Maximilian Ernestus, and Noah Dormann.
\newblock Stable-baselines3: Reliable reinforcement learning implementations.
\newblock {\em Journal of Machine Learning Research}, 22(268):1--8, 2021.

\bibitem[\protect\citeauthoryear{Shang \bgroup \em et al.\egroup }{2021}]{NEURIPS2021_ba3c5fe1}
Wenling Shang, Xiaofei Wang, Aravind Srinivas, Aravind Rajeswaran, Yang Gao, Pieter Abbeel, and Misha Laskin.
\newblock Reinforcement learning with latent flow.
\newblock In M.~Ranzato, A.~Beygelzimer, Y.~Dauphin, P.S. Liang, and J.~Wortman Vaughan, editors, {\em Advances in Neural Information Processing Systems}, volume~34, pages 22171--22183. Curran Associates, Inc., 2021.

\bibitem[\protect\citeauthoryear{Tan and Le}{2021}]{https://doi.org/10.48550/arxiv.2104.00298}
Mingxing Tan and Quoc~V. Le.
\newblock Efficientnetv2: Smaller models and faster training.
\newblock {\em ICML}, 2021.

\end{thebibliography}

\newpage
\section{Appendix}
\label{sec:appendix}
This appendix adds further implementation details to the networks used during the competition.
\subsection{Variational Auto-Encoder Architecture}
\label{subsec:variational_autoencoder_architecture_detail}
\begin{figure}[H]
    \centering
    \includegraphics[width=0.25\textwidth]{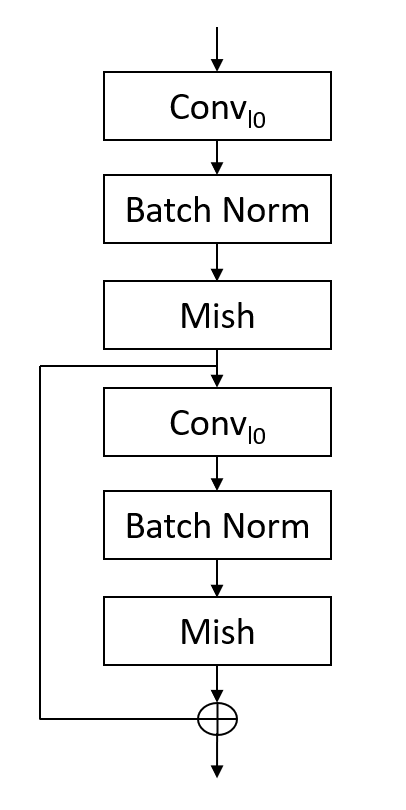}
    \caption{Detail of the modified convolutional blocks used.}
    \label{fig:vae2_convblocks}
\end{figure}

The baseline VAE provided in the competition was modified for the optical flow and edge images as described in section \ref{subsec:vae}. The convolutional blocks are summarised in Table \ref{tab:vae2_enc}, \ref{tab:vae2_dec} and a diagram detailing the layers of the blocks are shown in Figure \ref{fig:vae2_convblocks}. 

\begin{figure}[H]
    \centering
    \includegraphics[width=0.3\textwidth]{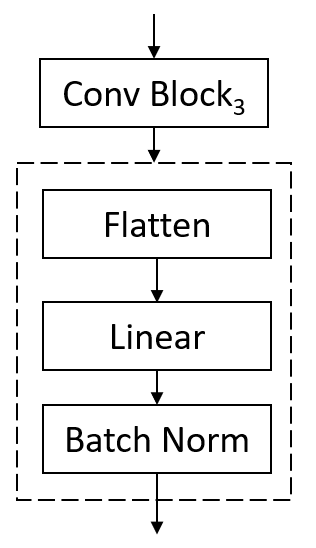}
    \caption{Detail of the modified fully connected blocks prior to re-parametrisation.}
    \label{fig:vae2_fcblocks}
\end{figure}

During training on the combined abstract input channels the network tended to become unstable which prevented a useful output. The fully connected layers were modified with normalisation to overcome the instability. The modification is shown in Figure \ref{fig:vae2_fcblocks}.

\begin{table}[H]
    \centering
    \begin{tabular}{l| c c c c c }
    \hline
    Stage & Block    & Channels & Kernel & Stride   & Pad \\
    \hline
    0     & Conv$_{00}$ & 4, 64    & 3     & 2         & 1 \\
    0     & Conv$_{01}$ & 64, 64   & 3     & -         & same \\
    1     & Conv$_{10}$ & 64, 128  & 3     & 2         & 1 \\
    1     & Conv$_{11}$ & 128, 128 & 3     & -         & same \\
    2     & Conv$_{20}$ & 128, 256 & 3     & 2         & 1 \\
    2     & Conv$_{21}$ & 256, 256 & 3     & -         & same \\
    3     & Conv$_{30}$ & 256, 512 & 3     & 2         & 1 \\
    3     & Conv$_{31}$ & 512, 512 & 3     & -         & same \\
    \hline
    \end{tabular}
    \caption{Modified VAE Encoder Architecture}\label{tab:vae2_enc}
\end{table}

\begin{table}[H]
 \centering
 \begin{tabular}{l| c c c c c }
     \hline
     Stage  & Block     & Channels    & Kernel & Stride & Pad \\
         \hline
         3      & Conv$_{31}$  & in\_ch, 512 & 3x3    & 2      & 1, 0    \\
         3      & Conv$_{30}$  & 512, 512    & 3      & 3      & 1 \\
         2      & Conv$_{21}$  & 512, 256    & 4      & 2      & 0, 1 \\
         2      & Conv$_{20}$  & 256, 256    & 3      & 1      & 1 \\
         1      & Conv$_{11}$  & 256, 128    & 3      & 2      & 2 \\
         1      & Conv$_{10}$  & 128, 128    & 3      & 2      & 2 \\
         0      & Conv$_{01}$  & 128, 64     & 3      & 2      & 2 \\
         0      & Conv$_{00}$  & 64, 4       & 3      & 1      & 1 \\
         \hline
     \end{tabular}
     \caption{Modified VAE Decoder Architecture}\label{tab:vae2_dec}
\end{table}

\newpage
\subsection{Segmentation Network Architectures}
\label{subsec:segmentation_network_architecture_detail}

As mentioned in section \ref{subsubsec:segmentation_model_architecture} the semantic segmentation model used was a EfficientNetV2 small encoder, paired with a FPN decoder. The EfficientNetV2 blocks are summarised in Table: \ref{tab:efficientnet_blocks}.

\begin{table}[H]
    \centering
    \begin{tabular}{c| c c c c}
    \hline
    Stage & Block & Stride & Filters & Layers \\
    \hline
    0  & Conv           & 2 & 24  & 1 \\
    1  & Fused-MBConv1  & 1 & 24  & 2 \\
    2  & Fused-MBConv4  & 2 & 48  & 4 \\
    3  & Fused-MBConv4  & 2 & 64  & 4 \\
    4  & MBConv4-SE0.25 & 2 & 128 & 6 \\
    5  & MBConv6-SE0.25 & 1 & 160 & 9 \\
    6  & MBConv6-SE0.25 & 2 & 256 & 15 \\
    \hline
    \end{tabular}
    \caption{EfficientNetV2-S Encoder Architecture Detail}
    \label{tab:efficientnet_blocks}
\end{table}

\begin{figure}[H]
    \centering
    \includegraphics[width=0.47\textwidth]{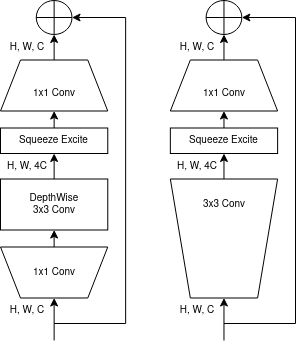}
    \caption{MBConv and Fused-MBConv}
    \label{fig:mb_conv_blocks}
\end{figure}

The difference between MBConv and Fused-MBConv are shown in Figure \ref{fig:mb_conv_blocks}.

The FPN decoder uses convolutional filters passed from the encoder at different resolutions. Filters summarised in Table: \ref{tab:efficientnet_filters_fpn},  input image size was (384, 512, 1).

\begin{table}[H]
    \centering
    \begin{tabular}{c | c | c | c}
    \hline
    Stage from table \ref{tab:efficientnet_blocks} & Resolution & Filters & Shape (h, w) \\
    \hline
    2 & 1/4  & 48  & 96x128  \\
    3 & 1/8  & 64  & 48x64   \\
    5 & 1/16 & 160 & 24x32   \\
    6 & 1/32 & 256 & 12x16   \\
    \hline
    \end{tabular}
    \caption{Encoders convolutional filters passed to FPN}
    \label{tab:efficientnet_filters_fpn}
\end{table}

\begin{figure}[H]
    \centering
    \includegraphics[width=0.47\textwidth]{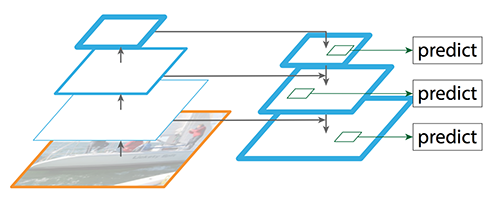}
    \caption{Feature Pyramid Network \protect\cite{fpn}}
    \label{fig:fpn}
\end{figure} 

\subsection{Classification Network Architecture Detail}
\label{subsec:classifier_network_architecture_detail}

As mentioned in section \ref{subsubsec:classification_model_architecture} the classification model was an MLP. The input feature vector of length 49152 was created by reshaping the lowest resolution convolutional filters from the EfficientNetV2 encoder, see last row in table \ref{tab:efficientnet_filters_fpn}. The model specifics used for our round 2 submission is detailed in table \ref{tab:classifier_layers_detail}.

\begin{table}[H]
    \centering
    \begin{tabular}{l| c c c}
        \hline
        Layer     & Input Size     & Output Size   & Activation \\
        \hline
        Input     & 49152  & 49152 & Tanh \\
        Hidden 1  & 49152  & 512   & ReLU \\
        Hidden 2  & 512    & 256   & ReLU \\
        Output    & 256    & 3     & Softmax \\
        \hline
    \end{tabular}
    \caption{MLP Layers used for location classification}
\label{tab:classifier_layers_detail}
\end{table}

\begin{table}[H]
    \centering
    \begin{tabular}{l c c| c c}
        \hline
        Track  & $fps$ & local kph & $fps$ & eval kph \\
        \hline
        Vegas North Road   &     &        & 5.1 & 92.54 \\
        Anglesey National & 7.0 & 103.97 &     &       \\
        Thruxton         & 7.0 & 147.12 &     &       \\
        \hline
    \end{tabular}
    \caption{Track times with location classifier, $fps$ frames per second, local or evaluation (AIcrowd) refers to simulator location}
    \label{tab:track_times_all}
\end{table}

\end{document}